\RequirePackage{fix-cm}
\documentclass[twocolumn]{svjour3}          % twocolumn
\smartqed  % flush right qed marks, e.g. at end of proof
\usepackage{graphicx,color}
\usepackage{bm,amsmath}
\usepackage{algorithm}
\usepackage{algpseudocode}
\usepackage{amsfonts}
\usepackage{graphicx}
\usepackage{enumitem}
\usepackage{mathrsfs}

\DeclareMathAlphabet\mathpzc{OT1}{pzc}{m}{it}
\let\mathcal=\mathpzc

\numberwithin{equation}{section}

\let\trueiiint=\iiint
\def\iiint{\mathop{\textstyle\trueiiint}\limits}
\def\intinfty{\int\limits_{\!\!-\infty\,\,}^{\,\,\infty\!\!}\kern-0.0em}
\def\iintinfty{\mathop{\int\!\!\int}\limits_{\!\!-\infty\,\,}^{\,\,\infty\!\!}\kern-0.0em}
\def\iiintinfty{\mathop{\int\!\!\int\!\!\int}\limits_{\!\!-\infty\,\,}^{\,\,\infty\!\!}\kern-0.0em}

\let\<=\langle
\let\>=\rangle
\let\^=\hat
\def\~#1{{\-ox{\sf#1}}}
\def\N{{\mathbb N}}

\let\-=\mathbf

\let\-=\mathbf

\usepackage{amsmath}
\usepackage{amssymb,bm}
\usepackage{graphicx}
\usepackage{color}
\usepackage{mathrsfs} 
\usepackage{enumerate}

\def\P{{\mathbb P}}

\def\N{{\mathcal{N}}}
\def\Nset{{\mathbb N}}

\def\_#1{{\underline{#1}}}

\let\<=\langle
\let\>=\rangle
\let\^=\hat
\def\~#1{{\mbox{\sf#1}}}
\let\'=\vec
\let\-=\mathbf
\newcommand{\co}{{\cal O}}

\newcommand{\bx}{{\mathbf x}}

\providecommand{\norm}[1]{\lVert#1\rVert}

\numberwithin{equation}{section}

\def\@#1{{\cal #1}}

 \journalname{Structural and Multidisciplinary Optimization BRIEF NOTE}
\begin{document}

\title{A derivative-free trust-region algorithm for reliability-based optimization\thanks{This work was
supported by the NSFC under grant number 11301337 and by the National Basic Research Program (973 Program) of China  under grant number 2014CB744900. }} 


\titlerunning{A DF-TR  algorithm  for RBO}        % if too long for running head

\author{Tian Gao \and
				Jinglai Li %etc.
}

\authorrunning{T. Gao and J. Li} % if too long for running head

\institute{Tian Gao \at
              Department of Mathematics,  
Shanghai Jiao Tong University, 800 Dongchuan Rd, Shanghai 200240, China.            
           \and
           J. Li \at
              Institute of Natural Sciences, Department of Mathematics, and 
the MOE Key Laboratory of Scientific and Engineering Computing, 
Shanghai Jiao Tong University, 800 Dongchuan Rd, Shanghai 200240, China.\\
\email{jinglaili@sjtu.edu.cn}
}

\date{Received: date / Accepted: date}

\maketitle

\begin{abstract}
In this note, we present a  derivative-free  trust-region (TR) algorithm for reliability based optimization (RBO) problems. 
The proposed algorithm consists of solving a set of  subproblems, in which simple 
surrogate models of the reliability constraints are constructed and used in solving the subproblems.  
Taking advantage of the special structure of the RBO problems, we employ a sample reweighting method to evaluate the failure probabilities, which  
constructs the surrogate for the reliability constraints by performing only a single full reliability evaluation in each iteration. 
With numerical experiments, we illustrate that the proposed algorithm
is competitive against existing methods. 

% \subclass{62F15 \and 65C05}
\end{abstract}
\section{Introduction}\label{sec:intro}

Reliability based optimization (RBO) problems, which optimize the system performance 
subject to the constraint that the system reliability satisfies a prescribed requirement, are 
an essential task in many engineering design problems~\cite{survey,aoues2010benchmark}. 
In a standard RBO problem, the reliability constraint is typically formulated as that the failure probability of the system is lower
than a threshold value,
and a very common class of RBO problems is to minimize a cost function
subject to the failure probability constraint:
    \begin{equation}
    \label{e:formulation}
    \min_{\bx \in D}   f(\bx), \quad
    s.t.\,\,  c(\-x) := \ln P(\-x) -\ln\theta\leq 0, 
    %\pr \{ g(\bx,\bm{\xi}) < 0 \} < \epsilon.
    \end{equation}
where $\bx$ is the design parameter, $D$ is the design space, $f(\cdot)$ is the cost function, $P(\bx)$ is the failure probability associated with design $\bx$ and $\theta$ is the failure probability threshold.  
%and $\bm{\xi}$ is random variable representing the uncertainty in the systems.
In practice, the cost function is often deterministic and easy to evaluate, 
 while computing the probabilistic constraint is much more costly as it requires
expensive Monte Carlo (MC) simulations.
%We will discuss how to compute the failure probability in next section, and here we simply regard it as a computationally intensive function.  

In this work we  consider  the so-called double loop~(DL) RBO methods, where an inner loop estimating the failure probability is nested in
 the outer loop solving the optimization problem~\cite{aoues2010benchmark,survey,eldred2002formulations}, 
 and so other methods, such as the single loop and the decoupling algorithms~\cite{survey} are not in our scope.   
The DL methods only require to evaluate the limit state function of the underlying system, which makes it particularly convenient for problems with black-box models.
%The DL approaches are often computationally intensive. 
The computational burden of the DL methods arises from both the inner and the outer loops. 
Namely, the total computational cost depends on the number of reliability (failure probability) evaluations required  and the cost for performing
each single failure probability evaluation.
This work aims to address the former:  to solve the RBO problems with a small number of reliability evaluations. 
A difficulty here is that,  due to the use of MC simulations, it is very difficult to obtain the derivatives of  the reliability
constraints.
%Several works (e.g. \cite{lee2011sampling,wang2014maximum,dubourg2014meta})
One way to  alleviate the difficulty is to perform stochastic sensitivity analysis with the so-called score functions (SF)~\cite{Rahman2009278,rubinstein1993discrete}.
Here we consider an alternative type of methods, known as the derivative-free (DF) trust-region (TR) algorithms~\cite{DFO}, developed to solve problems whose derivatives are difficult to obtain. 
Loosely speaking, the DF-TR methods consist of solving a set of TR subproblems in which surrogate models of the  objective
and/or the constraint functions are constructed and used in solving the subproblems.    
%To the best of our knowledge, no DF-TR based methods have been developed for the RBO problems. 
{ The main contribution o the work is two-fold. 
First we present a DF-TR algorithm specifically designed for the RBO problems, which does not require the knowledge of the derivative 
information of the objective and 
the constraint functions. 
Note that the computational cost associated with the DF-TR algorithm poses a  challenge here, as
constructing a surrogate model with regression or interpolation requires to repeatedly evaluate the reliability constraints, which is highly expensive. 
Thus our second contribution is to employ a sampling reweighting method, which only uses
 a \emph{single full reliability evaluation} to construct the surrogates in each TR iteration.}
With a numerical example, we illustrate that the DF-TR algorithm can be a competitive alternative to the score-function based methods. 

The paper is organized as follows. We present our DF-TR algorithm for RBO problems in Section~\ref{sec:algorithm}.
We describe the evaluation of reliability constraints in Section~\ref{sec:ce}.  Finally we provide a benchmark example to demonstrate the performance of the proposed algorithm in 
Section~\ref{sec:examples}.  % Finally we offer some concluding remarks in Section~\ref{sec:conclusions}.

\section{The DF-TR algorithm for RBO problems}\label{sec:algorithm}
%Firstly we discuss about the formulation of the problem and the optimization procedure, set up a framework for solving it. Details about the determination of %the trust-region are presented in the second subsection, along with the algorithm for surrogate model construction. Lastly we will prove convergence of the %algorithm under the derivative-free trust-region scheme.

%%%%%%%%%%%%%%%%%%%%%%%%%%%%%%%%%%%%%%%%%

%\subsection{The derivative-free trust-region algorithm}

A natural idea to solve the RBO problem~\eqref{e:formulation} is to construct a computationally efficient surrogate for the constraint $c(\-x)$, and then solve the optimization problem subject to the surrogate constraint. 
The TR methods provide a rigorous formulation of this surrogate based approach. 
The TR methods start from an initial point $\bx_0$ and finds a critical point by computing a series of intermediate points $\{\-x_k\}_{k\in \Nset}$. 
Specifically, suppose the current point is $\-x_k$, and to compute the next point, the algorithms solve a TR subproblem in which
surrogates of the objective function and the constraints are constructed and used in a neighborhood of $\-x_k$. 
This neighborhood of $\-x_k$ is known as the trust-region and the size of it is adjusted in a way that the surrogate models are sufficiently accurate in it.
In our problem,  the objective function is of simple form, and we only need to construct the surrogate for the constraint function. 
As a result, in our RBO problems, the TR sub-problem at iteration $k$ becomes,
    \begin{equation}
    %\begin{aligned}
    \min_{\bx\in D} f(\bx), \quad %\notag\\
    \mathrm{s.t.} \,  s_k(\bx) \leq 0 \, \mathrm{and}\, \|\bx-\bx_k\|\leq \rho_k, \label{e:sub}
    \end{equation}
%    \end{equation}
where $s_k(\bx)$ is the surrogate model of $c(\bx)$, and  $\rho_k$ is the radius of the TR of $\bx_k$. 
In what follows we use the notation: $\co(\-x_c,\rho) = \{\bx | \|\bx-\-x_c\|\leq \rho\}$. 
Before discussing the construction of the surrogate models, we first present our main algorithm for solving the RBO problems: 
%We describe the complete procedure in Algorithm~\ref{alg:opt}:
   
  %  \label{alg:opt}
    \begin{algorithmic}[1]

    \Require
$f(\bx)$,  $c(\bx)$, $\bx_0$,  $\rho_{0}$,  $\rho_{\min}$, 
    $\omega^+$,  $\omega^-$,   $\delta$, $\epsilon^*$,  $M$.
    \Ensure
    Solution $\bx_\mathrm{opt}$;
    %Function value $f^\ast$;
    %Failure probability $\pr^\ast$.
    % Algorithm
    
   \State Outer:$=1$; $k:=0$;
    \While {Outer$=1$}
       \State  Inner$:=1$;
			\While{Inner$=1$}
       \State $[s_k(\bx),\rho_k] := \mathrm{SurrConstr}(\bx_k,\rho_k,\epsilon^*,\omega^-,M)$;
        \State $\bx_{k+1}:= \arg\min_{\bx\in\co(\bx_k,\rho_k)} f(\bx)$, s.t. $s_k(\bx) \leq 0$;
        \If {$c(\bx_{k+1}) < 0$ } \label{ln:cond}
            \State Inner$:=0$;
						\State $\rho_k = \omega^+\rho_k$;
						\Else
						\State $\rho_k := \omega^-\rho_k$;
             
						%\State  goto line 3;
        \EndIf
				\EndWhile
        %\State $//$ construct model $m_{k+1}(\bx)$ in trust-region $\co(\bx_{k+1}, \rho_{k+1})$;
        \If {$\norm{\bx_{k+1}-\bx_k} < \rho_k$ or $\|f(\bx_{k+1})-f(\bx_k)\| \leq \delta$ or $\rho_k<\rho_{\min}$}
            \State Outer$:=0$; %exit the \textbf{for} loop;
						\Else
						\State $\rho_{k+1} := \rho_k$;
						\State $k:=k+1$;
        \EndIf
    \EndWhile
\State    $\bx_\mathrm{opt}:= \bx_k$;

		{\bf Alg. 1:  The DF-TR RBO algorithm}	
    \end{algorithmic}
	%	\medskip
   % \end{algorithm}
%%%%%%%%%%%%%%%%%%%%%%%%%%%%%%%%%%%%%%%%%
%%%%%%%%%%%%%%%%%%%%%%%%%%%%%%%%%%%%%%%%%

A key step in a TR algorithm is to adjust the radius of the TR in each step. In this respect our algorithm follows the procedure given in \cite{NOWPAC}, 
but only adjusts the radius according to the constraint function (\cite{NOWPAC} adjusts it based on both the objective and the constraint functions). 
{Here $\rho_0$ is the initial TR radius and $\omega^+$ and $\omega^-$ are the TR expansion and contraction constants respectively.  
The TR subproblem~\eqref{e:sub} can be solved with any usual constrained optimization technique, and in this work 
we choose to use the sequential quadratic programming (SQP) method. 
 The algorithm terminates when one of the following three conditions is satisfied: %the number of iterations exceeds a prescribed maximum value; 
 $\bx_{k+1}$ is an inner point of $\co(\bx_k,\rho_k)$, 
 the difference between $f(\bx_k)$ and $f(\bx_{k+1})$ is below a prescribed threshold $\delta$,
or the radius is smaller than a prescribed minimal value $\rho_{\min}$. 
Moreover, $\epsilon^*$ is the error bound for the surrogate models,
and $M$ is the number of points used to construct the surrogate models. }

We now discuss the construction of the surrogates, which is a critical step in Algorithm~1. 
In the DF framework, one first writes the surrogate model as a linear combination of a set of basis functions
 namely, 
    \begin{equation}\label{e:mx}
    s(\bx) = \sum_{l=1}^L a_l b_l(\-x),
    \end{equation}
where  $\{b_l(\bx)\}_{l=1}^L$ are a set of basis functions and $\-a=(a_1,...,a_L)^T$ is the vector collecting all the coefficients, 
 and then determines the coefficients $\-a$
with either regression or interpolation. 
We choose to use the popular \emph{quadratic polynomials} surrogates, 
while noting that the proposed algorithm does not depend on any particular type of surrogates.
% The coefficients $\-a$ in Eq.~\eqref{e:mx} can be determined by either interpolation or regression. 

%The surrogate construction procedure highlights a difference between our algorithm and the standard DF-TR methods. 
In the standard DF-TR algorithms, the surrogate models are required to be fully linear or quadratic~\cite{DFO}. 
Imposing such conditions is very difficult in RBO problems as the failure probability is evaluated with sampling methods.
Thus here we simply require that the error between the surrogate and the true constraint function is bounded in the TR: 
for a given fixed $\epsilon > 0$ and a TR $\co(\-x_c,\rho)$,  
    $|s(\-x)-c(\-x)|\leq\epsilon$ for any $\bx\in \co(\-x_c,\rho)$.
Now, we propose a scheme to construct TR surrogate with a bounded error, described as:
    
    \centerline{\noindent\bf Alg.~2: $[s(\cdot),\rho]=\mathrm{SurrConstr}(\-x_c,\rho_\mathrm{max},\epsilon^*,\omega, M)$}
    
    \begin{algorithmic}[1]
    \State let $\rho:=\rho_\mathrm{max}$; LOOP:$=1$;
    \While{LOOP$=1$}

         \State randomly generate $M-1$ points in $\co(\bx,\rho)$: $\{\bx_m \in \co(\bx, \rho)\}_{m = 1}^{M-1}$; \label{line:start}
				\State let $\bx_M=\-x_c$;
         \State  evaluate the constraint function $y_m=c(\bx_m)$ for $m=1...M$. %and denote the results as $Y$;
       
         \State  compute $s(\bx)$ using data  set $\{(\bx_m, y_m)\}_{m=1}^M$; 
          \State estimate the approximation error bound $\epsilon$ of $s(\cdot)$ with leave-one-out cross validation; \label{line:error}
          \If  {$\epsilon<\epsilon^*$} 
					\State LOOP$:=0$;
					\Else
					\State $\rho := \omega\rho$;
									\EndIf

    \EndWhile
    \State \Return $s(\cdot)$ and $\rho$. \label{line:end}

    \end{algorithmic}

Simply put, the algorithm constructs the quadratic regression and examines whether the resulting surrogate
satisfies the error bound condition; %if the maximum iteration number is reached and the assumption has yet been satisfied,
if not, the algorithm contracts the TR and repeats.  
In Line \ref{line:error}, we estimate the approximation error with the leave-one-out cross validation  method. Namely, 
let $X = \{\bx_1,...,\bx_M\}$ and $Y=\{y_1,...,y_M\}$ with $y_m=c(\bx_m)$ for $m=1...M$.
Let $X^{m}_- = \{\bx_1,...,\bx_{m-1},\bx_{m+1},...,\bx_M\}$ and $Y^m_-=\{y_1, ...,$ $y_{m-1},y_{m+1},...,y_M\}$.
Let $s^m(\bx)$ be the surrogate model based on data $(X^m_-,Y^m_-)$ and the approximation error $\epsilon$ is estimated by
$\epsilon = \max\{ |c(\bx_m)-s^m(\bx_m)|\}_{m=1}^M$.
Apparently, to construct the surrogate, we need to evaluate the reliability constraint at a rather large number of design points, which can be computationally demanding. 
However, as will be shown in the next section, we apply a sample reweighting strategy, which 
allows us to obtain the values of the constraint at all the design points by only performing a full sampling based reliability evaluation at $\-x_c$.
Thus the computational cost is significantly reduced.  
{We also note that, another way to improve the efficiency for evaluating the reliability constraint is to use low-cost surrogate models for the limit state
function, but in many practical problems (e.g. the source of uncertainty is modeled by a random process),  constructing such surrogates itself can be a very challenging task. }

%%%%%%%%%%%%%%%%%%%%%%%%%%%%%%%%%%%%%%%%%
%%%%%%%%%%%%%%%%%%%%%%%%%%%%%%%%%%%%%%%%%

\section{The sample reweighting method}\label{sec:ce}
In this section, we discuss the evaluation of the reliability constraint $c(\-x)$,
or equivalently, the failure probability $P(\bx)$.  
Let  $\-z$ be a $d_z$-dimensional random variable with distribution $q(\-z)$, representing the uncertainty in a system. 
The system reliability is described by the limit state function  $g(\-z)$, and, namely, 
 the event of failure is defined as  $g(\-z)<0$.  
Following the formulations in \cite{Rahman2009278}, we assume that the distribution of $\-z$ depends 
on the design parameter $\-x$, i.e., $q(\-z;\-x)$, while the limit state function $g(\-z)$ is independent of $\bx$.
%\begin{equation}
%P(\-x) =\int  I(\-z)q(\-z;\-x) d\-z,\label{e:pf2}
%\end{equation}
%where $q(\-z;\-x)$ is the PDF of $\-z$ depending on the design parameter $\-x$. 
As a result the failure probability is
\begin{equation}
P(\-x) = \P(g(\-z)<0) = \int_{\-z\in R^{d_z}} I(\-z) q(\-z,\bx) d\-z,\label{e:pf}
\end{equation}
where  $I(\-z)$ is an indicator function:
\begin{equation}
I(\-z) = \big\{ \begin{array}{ll}
         1 &\quad \mbox{if $g(\-z) < 0$},\\
         0 & \quad \mbox{if $g(\-z) \geq 0$}. \end{array} %\right\}
\end{equation}
%In what follows we shall omit the integration domain when it is simply $R^{d_z}$.
$P(\bx)$ can be computed with the MC estimation:
\begin{equation}
  \hat{P}_\mathrm{MC}= \frac{1}{N}\sum^N_{n=1}{I}(\-z^{(n)}),
\end{equation}
with samples $\-z^{(1)},...,\-z^{(N)}$ drawn from $q(\-z;\bx)$.

Recall that in Algorithm~1, we need to evaluate the failure probability at a number of design points in the TR to construct the surrogate function. 
Since each evaluation requires a full MC sampling procedure, the total computational cost can be very high. 
To improve the efficiency, we present a sample reweighting approach, which allows one to obtain the failure probability values at all design points
with one full MC based failure probability evaluation. 
Suppose we have performed a MC estimation of the failure probability at the center of the TR, $\-x_c$, obtaining a set of samples from 
$q(\-z;\bx_c)$:
$\{(\-z^{(n)},g(\-z^{(n)}))\}_{n=1}^N$.  
 For any point $\-x$ in the TR, we can write $P(\-x)$ as, 
\begin{equation}
P(\-x) = \int I (\-z) q(\-z;\-x) d\-z 
=\int I(\-z)  r(\-z) q(\-z;\bx_c) d\-z,
\end{equation}
where $r(\-z) = q(\-z;\-x)/q(\-z;\-x_c)$.
It follows immediately that $P(\-x)$ can be estimated as
\begin{equation}
\hat{P}(\-x) = \frac1N\sum_{n=1}^N I(\-z^{(n)}) r(\-z^{(n)}),\end{equation}
i.e., by simply assigning new weights $r(\-z)$ to the samples generated in the evaluation of $P(\-x_c)$. 
Note that, in this method, only the computation of $P(\-x_c)$ involves the evaluations of the limit state function $g(\cdot)$,  which is referred to as \emph{a full reliability evaluation}.  
{This method uses the same formulation as importance sampling (IS), 
but it differs from a standard IS as its purpose is not to reduce the sampling variance, but to reuse the samples. } 
We use this approach to construct the surrogates in Alg.~2.

\section{A benchmark example}\label{sec:examples}

As an illustrating example, we consider a cantilever beam problem, with width $W$, height $T$, length $L$,
and subject to transverse load $Y$ and horizontal load $X$.
This is a well adopted benchmark problem in optimization under uncertainty~\cite{eldred2002formulations}, where the system failure
is defined as the maximum deflection exceeding a threshold value:
\begin{equation}
g =D_o-\frac{4L^3}{EWT}\sqrt{\left(\frac{Y}{T^2}\right)^2+\left(\frac{X}{W^2}\right)^2}.
\end{equation}
Here $D_o$ is the deflection threshold and $E$ is the Young's modulus. 
In this example we assume the beam length $L$ is fixed to be $100$ and $D_0=6$.
 The random variables are:  the elastic modulus $E \sim \N(29 \times 10^6,(1.45 \times 10^6)^2)$,  external loads 
$X \sim \N(500, 25^2)$ and $Y \sim \N(500, 25^2)$, and the actual beam width $W\sim \N(w,\sigma^2)$ and height $T\sim \N(t,\sigma^2)$, respectively. 
The mean width $w$ and the mean height $t$ are design variables, and 
our goal is to minimize the construction cost 
$f(w,t) = w t$, %\label{e:beamopt}
subject to that the associated failure probability is smaller than $\theta=0.1$. 
In the numerical tests, we solve the problem with $\sigma=10^{-1}$ and $\sigma=10^{-2}$.

{For comparison, we solve the problem with three methods:
the DF-TR with reweighting (denoted by DF-TR-R),
the DF-TR method without reweighting (denoted by DF-TR),
 and a standard active set method, where the gradients are computed 
with the SF method (denoted by SF). }
The algorithm parameter values of the DF-TR and DF-TR-R algorithms are given in Table~\ref{tb:P1params}.
In the MC simulations of all the methods, we use two samples sizes $N=10^{4}$ and $N=10^5$. 
Since all the methods are subject to random errors, to take that into account,
we repeatedly solve the problem with all the three methods 100 times
and summarize the results in Table~\ref{tb:P2results}. 
{Specifically, we compare the average errors of the obtained solutions (compared to a benchmark solution computed by the SF method with $5\times10^6$ samples),
and the average number of function evaluations. }
We see from the results that 
in all the test cases, the DF-TR-R algorithm outperforms the SF based method,
in terms of both average errors and the number of full reliability evaluations.
 This suggests that the proposed DF-TR-R algorithm can be more robust and efficient than the SF method for small sample size. 
In the comparison of the two DF-TR algorithms, we can see that, both algorithms 
 yield comparable results in terms of accuracy, while the DF-TR-R algorithm uses significantly less 
 full reliability evaluations than the algorithm without reweighting. 
We note that more numerical tests are needed to have a conclusive performance comparison of the methods. 
Nevertheless, the results suggest that the DF-TR-R algorithm provides an efficient and easy-to-use alternative to the SF based methods. 

\begin{table}[!htb]
\center
\begin{tabular}{c c c c c c c}
\hline
 $ \rho_{0} $ &$\rho_{\min}$& $  \epsilon^* $ & $ \omega^-$ & $ \omega^+$  & $ M$&$\delta$\\
\hline
 0.1 &$10^{-6}$ &$0.1\theta$  & 0.9&1.1 &20&$10^{-4}$ \\
%& $2.90\times 10^{-3}$ & $3.98\times 10^{-3}$ & $1.62\times 10^{-2}$ & $4.16\times 10^{-4}$ & $1.02\times 10^{-2}$ \\
\hline
\end{tabular}
\caption{The parameter values of the DF-TR algorithm.}
\label{tb:P1params}
\kern-4\medskipamount
\end{table}

\begin{table}[!htb]
\center
\ {
\begin{tabular}{llccc}
\hline %
$\sigma$&{$N$}&method& avg error&   full evals\\
\hline %
 & & SF & 0.079  & 142 \\ \cline {3 -5} 
 $10^{-1}$& $10^4$  &  DF-TR & 0.025  & 620 \\ \cline {3 -5} 
& &  DF-TR-R & 0.0273  & 49 \\ \cline {2 -5} 
 &   & SF &  0.063  & 140 \\ \cline {3 -5} 
 $10^{-1}$& $10^5$&  DF-TR & 0.021  & 380 \\ \cline {3 -5} 
 & &  DF-TR-R & 0.0217  & 28 \\ \hline\hline
  &   & SF & 0.0334  & 114 \\ \cline {3 -5} 
$10^{-2}$  & $10^4$&  DF-TR & 0.015  & 420 \\ \cline {3 -5} 
& &  DF-TR-R & 0.0201  & 29 \\ \cline {2 -5} 
 &   & SF & 0.031  & 126 \\ \cline {3 -5} 
 $10^{-2}$& $10^5$&  DF-TR & 0.011  & 320 \\ \cline {3 -5} 
 & &  DF-TR-R & 0.0165  & 18 \\ \hline
\end{tabular}
}
\caption{Performance comparison of the three methods.}
\label{tb:P2results}
\kern-5\medskipamount
\end{table}

%\begin{figure}[!htb]
%\centerline{\includegraphics[width=.35\textwidth]{figs/P2_path}}
%\centerline{\includegraphics[width=.35\textwidth]{figs/P2_fval}}
%\caption{Left: the path from the initial guess to the final solution (the solid line is the boundary of the feasible region, also computed with the CE method).  
%Right: the function value plotted against the number of iterations and the dashed line
%shows the cost associated with the exact solution.}\label{fig:P3results}
%\end{figure}

\section{Conclusions}\label{sec:conclusions}

In summary, we present a DF-TR algorithm to solve the RBO problems without using the gradients of the reliability constraints. 
A sample reweighting method is employed so that the TR surrogate can be obtained by performing a single 
full reliability evaluation. 
Due to space limitation we only present a simple benchmark example, and 
applications of the method to some real-world design problems will be reported in a future work.  
{Moreover, we note that in general the design parameters $\bx$
could also affect the limit state function itself,
and in this case the sample reweighting method does not apply directly. 
We hope to address this issue in future studies. }

\bibliographystyle{spmpsci}      % mathematics and physical sciences
\bibliography{rbo}



\end{document}